\documentstyle[newpasp,twoside,epsf]{article}
\def\delv{\Delta v}
\def\kms{\, \rm{km}\,  \rm{s}^{-1}}
\def\cm2{{\, \rm{cm}}^{-2}}

\begin{document}
\title{Disks at High Redshift: Interactions, Mergers, and Starbursts}
\author{Rachel S. Somerville}
\affil{Institute of Astronomy, University of Cambridge, Madingley Rd., 
Cambridge, UK CB3 0HA}

\begin{abstract}
Do disk galaxies exist at redshifts much greater than unity, and how
might they look different from local disks? How does the morphological
mix of galaxies change with redshift?  What can we learn from current
observations about the properties of high redshift galactic disks?  I
present theoretical predictions based on semi-analytic hierarchical
models, focussing on the role played by interactions, mergers, and
starbursts in determining the observable properties of disk galaxies
at high redshift, and discuss the interpretation of high redshift
observations of possible proto-disks (damped Lyman-$\alpha$ systems
and Lyman-break galaxies) in light of these predictions.
\end{abstract}

\section{Introduction}
Very little is known about the properties of disk galaxies at
redshifts larger than about unity; even whether they exist at all.
The high redshift universe is a rather dangerous place for galactic
disks, which are relatively fragile objects. Mergers are more frequent
at early times, and the characteristic gas fractions and internal
densities of early objects are high. Are violent mergers so frequent
at high redshift that they destroy most or all disks? Are high
redshift disks more or less susceptible to bar instabilities than
local disks? 

Before we begin to discuss theoretical answers to these questions,
based on the currently accepted paradigm of disk formation within a
hierarchical universe, let us address a semantic and practical
question --- what do we mean by a ``disk'' galaxy, and how do we know
one when we see it? The defining feature of a ``true'' disk is
fundamentally kinematical: disks are objects that are flattened and
supported by rotation. In the nearby universe, we can obtain direct
observational measures of galaxy kinematics and (in principle)
determine unambiguously whether an object is a disk or not. However,
these observations become increasingly difficult as one pushes to high
redshift.  A small sample of disk galaxies with observed rotation
curves extends out to a redshift of about unity (the current record
holder is at $z=1.3$; N. Vogt, private communication). For the moment,
and probably for the near future, if we wish to identify disks at
still higher redshifts, we will be forced to rely on more indirect
observational evidence.

For example, we know that in the local universe, if we select galaxies
by their morphology (spirals and irregulars), or the shape of their
light profiles (exponential), or those that have blue colors or
strong emission lines, we will end up with a sample that consists
mostly of ``true'' disks. However, by some of these conditions, some
interlopers (starbursts?) might creep in, and some ``true'' disk
galaxies (early-type spirals and lenticulars) might be left out. It is
unclear whether these empirical associations will remain useful at
high redshift. None the less, I will discuss what they might tell us
about the $z\sim3$ Lyman-break galaxy population (e.g. Steidel et
al. 1999). Another intriguing kind of observation is the kinematic
data from unsaturated, low-ionization-state metal lines in high column
density quasar absorption systems (damped Lyman-$\alpha$ systems
(DLAS); Prochaska \& Wolfe 1997, 1998). These absorption profiles may
provide kinematic information about gaseous disks at high redshift;
however, as I discuss in Section~3, interpreting these data may be
complicated.

\section{Mergers and Morphological Evolution of Disks}
\begin{figure}
\plottwo{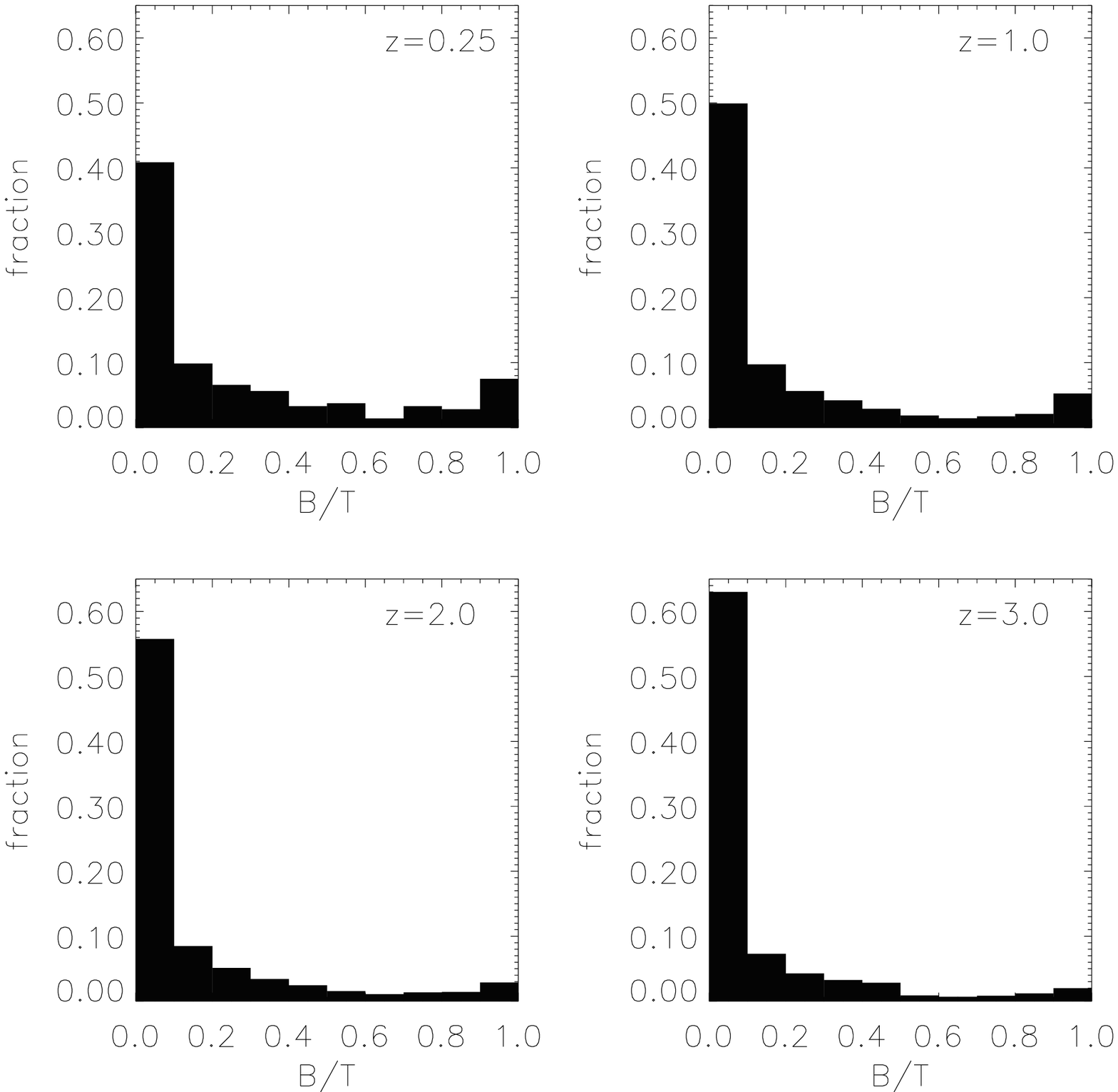}{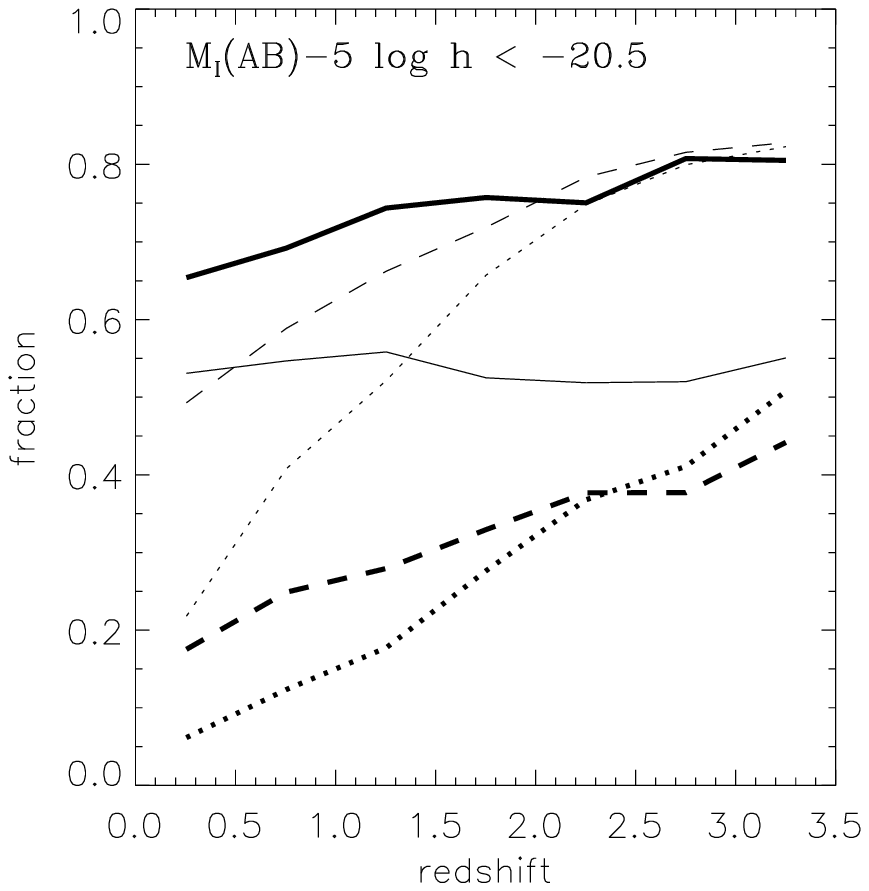}
\caption{Redshift evolution of the morphological mix of galaxies
with rest-frame absolute magnitude (AB) $M_I < -20.5 + 5
\log h_{100}$, selected from a semi-analytic model.
Left: The distribution of bulge-to-total stellar mass ratio. Note the
gradual depletion of very ``disky'' galaxies ($B/T
\la 0.1$) from $z=3$ to the present, and the corresponding build-up of
very ``bulgey'' ($B/T \ga 0.9$) galaxies.  Right: The fraction of
various types of disk galaxies as a function of redshift.  The bold
solid line shows the fraction of all selected galaxies that have at
least half of their stellar mass in a disk component. The thin solid
line shows the fraction of these galaxies that may be unstable to bar
formation (see text). Dotted lines show the fraction of the bright
disk galaxies that have experienced a merger within the past 0.5 Gyr
(bold lines show the fraction that have experienced a major merger;
thin lines any sort of merger). Dashed lines show the fraction that
have experienced a merger within a time shorter than 10 times the
present internal dynamical time of the galaxy (again, bold lines show
major mergers, thin lines all mergers).  These are objects that will
almost certainly appear morphologically disturbed, and may exhibit
experiencing starburst and/or AGN activity.  }
\label{fig:morph}
\end{figure}
In this section I discuss some predictions from a semi-analytic model
representing the main features of the currently standard paradigm of
disk formation. In this paradigm, dark matter halos acquire angular
momentum from tidal torques at early times. Hot gas contained in these
halos is assumed to initially possess the same specific angular
momentum as the halo, and to conserve it as it cools and collapses to
form a rotationally supported, exponential disk. If the resulting disk
is locally unstable (Toomre $Q$), it will quickly fragment into many
dense knots. If it is globally unstable (Toomre $X$), it will form a
bar and perhaps a bulge. Subsequently, these disks may be destroyed by
major mergers (defined here as an encounter between objects with a
mass ratio of 0.3 or greater). Morphological evolution toward
`earlier' type disks can also be driven by minor mergers, which can
drive gas inflows, leading to nuclear starbursts that may build
central bulge-like structures (Mihos \& Hernquist 1994a).

Fig.~\ref{fig:morph}a shows the distribution of the bulge-to-total
ratio of stellar mass at various redshifts for galaxies produced by
such a model \footnote{The model shown is based on the models
described in Somerville \& Primack (1999), with more realistic disk
formation included using an approach similar to that of Mo, Mao \&
White (1998).}. As time progresses, disky galaxies are gradually
transformed into more bulge-dominated galaxies, and the distribution
changes shape noticeably --- however, the net effect is not
dramatic. Recall that although disks are constantly being destroyed by
mergers, hot gas is also continuously cooling and forming new
disks. Thus an individual galaxy can move in either direction along
the Hubble sequence over time\footnote{Note that I have selected model
galaxies with \emph{rest-frame absolute} I-band magnitudes (AB)
brighter than $-20.5+5 \log h_{100}$ (where $h_{100}
\equiv H_0/(100$ km/s/Mpc)). Had I instead selected galaxies according
to their observed frame
\emph{apparent} magnitude, as in a flux-limited survey,
the trend would have been washed out by the inclusion of more
intrinsically faint galaxies (which tend to have low $B/T$ values) at
low redshifts. This illustrates the importance of sample selection in
assessing observational trends.}.

Fig~\ref{fig:morph} shows the fraction of bright galaxies (rest-frame
$M_I(AB) < -20.5+5 \log h_{100}$) that are disk-dominated ($B/T <
0.5$) as a function of redshift. The overall fraction of disk galaxies
defined in this way increases only moderately with redshift, from 65
percent at $z=0.25$ to 80 percent at $z=3$. This summarizes the
interplay between disk destruction (and spheroid production) by major
mergers and growth of new disks by gas cooling and star formation. We
have not yet considered the fact that some of these disks might be so
distorted that they would not satisfy our usual morphological
criteria, or that they might be spontaneously bar-unstable, which
could lead to secular formation of a bulge.  Therefore I have also
shown the fraction of these galaxies that are formally unstable to bar
formation, according to the condition $\epsilon_m < 0.9$, as defined
by Efstathiou, Lake, \& Negroponte (1982) and used in the disk models
of Mo et al. (1998). This is in reasonable accord with observations at
low redshift, which suggest that one-third to one-half of all galaxies
are barred, especially since I have neglected the stabilizing influence
of pre-existing bulges, so the fraction of unstable disks may have
been over-estimated, particularly at lower redshifts. The fraction of
bar-unstable galaxies is predicted to be approximately constant with
redshift\footnote{Again, the sample selection is important. In a
flux-limited sample, there is a decrease in the fraction of
bar-unstable disks with increasing redshift, which again is due to the
inclusion of more small galaxies (which tend to be bar-unstable) at
lower redshifts.}.  In addition, the fraction of disk galaxies that
have experienced a recent merger is shown as a function of redshift. I
use two definitions of ``recent''; in one, a fixed timescale of 0.5
Gyr, and another, a time span of 10 times the internal dynamical time
of the galaxy. Based on experience with N-body simulations, these are
galaxies that are likely to show morphological signatures of
disturbance, as well as starburst and possibly AGN activity. I return
to the problem of quantifying these effects in the last section of
this paper. We see that the fraction of disks that suffered recent
major mergers varies from 5--20 percent at redshift zero to about 50
percent at redshift three, while the fraction of disks that have
experienced recent minor mergers changes from 20--50 percent at
redshift zero, to 80 percent at redshift three.

So, with respect to the question that I introduced in the opening
paragraph, the theoretical prediction is that a large fraction of high
redshift galaxies should have substantial disk components, but many or
most of these disks should carry the signatures of recent mergers. I
turn now to the observations: do they support or refute this
prediction?

\section{Are the DLAS proto-disks?}
\label{sec:dlas}
Damped Lyman-$\alpha$ systems (DLAS) have traditionally been
interpreted as the high-redshift analogs of present day spiral disks
(e.g. Lanzetta, Wolfe, \& Turnshek 1995). The velocity profiles
obtained from unsaturated, low-ionization-state metal lines are
asymmetric and ``edge-leading'' (the strongest feature tends to occur
at one edge of the system), and the distribution of velocity widths
extends to large values of $\delv \sim 200-300
\kms$. Prochaska \& Wolfe (1997, 1998)
showed that their data was well fit by a ``thick disk'' model, in
which each DLAS corresponds to a rotationally supported disk, with a
scale height rather larger than typical nearby spirals, and a fairly
rapid rotation speed of $200
\kms$. This would imply that disks nearly as massive as
present day $L_*$ spirals are in place at redshifts of 2--3, which is
at odds with the predictions of hierarchical models. Indeed, the
$\delv$ distributions obtained from the disk formation models of
Kauffmann (1996) and Mo et al. (1998) are dramatically inconsistent
with the observations, as emphasized by Prochaska \& Wolfe (1998).

\begin{figure}
\plottwo{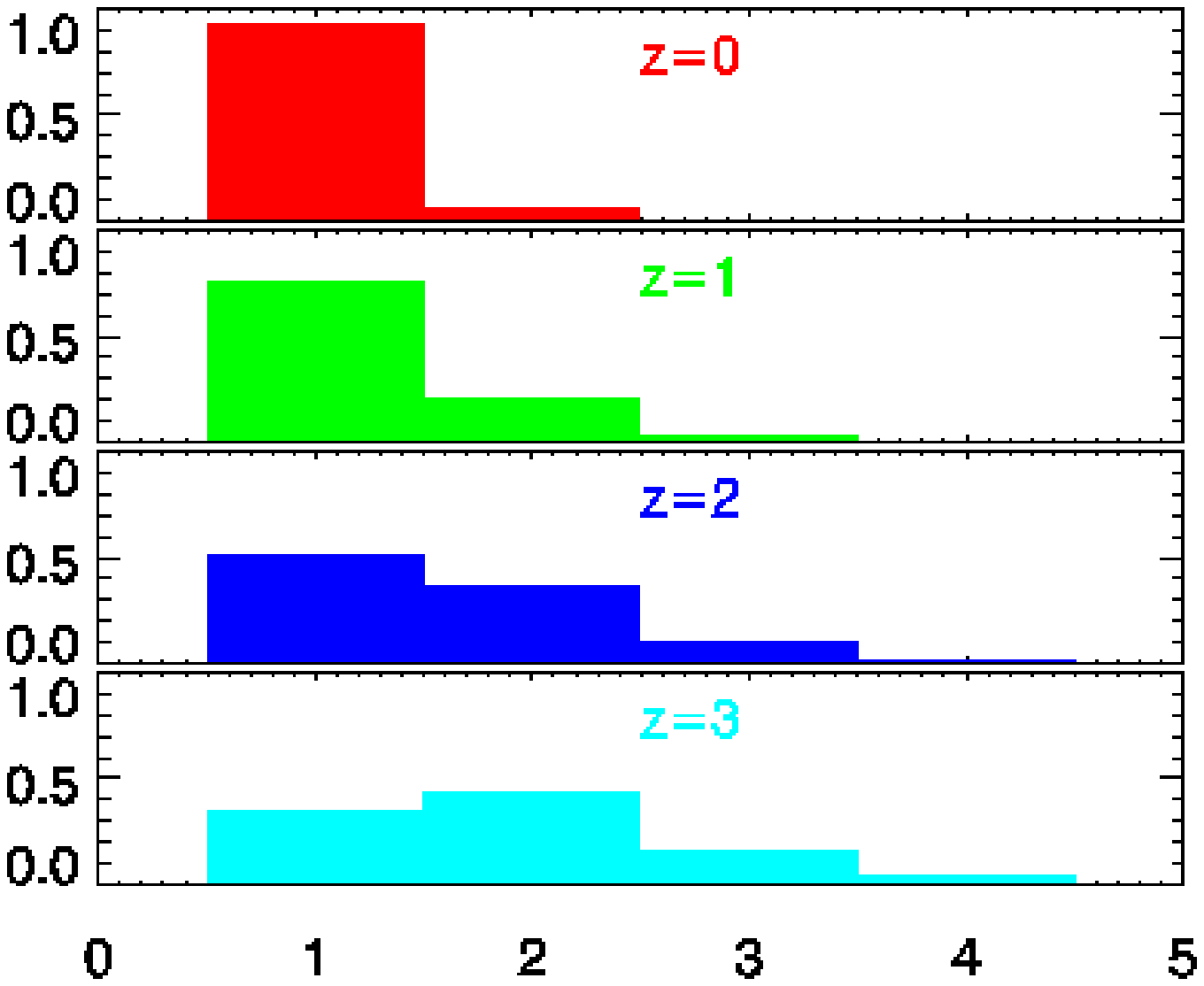}{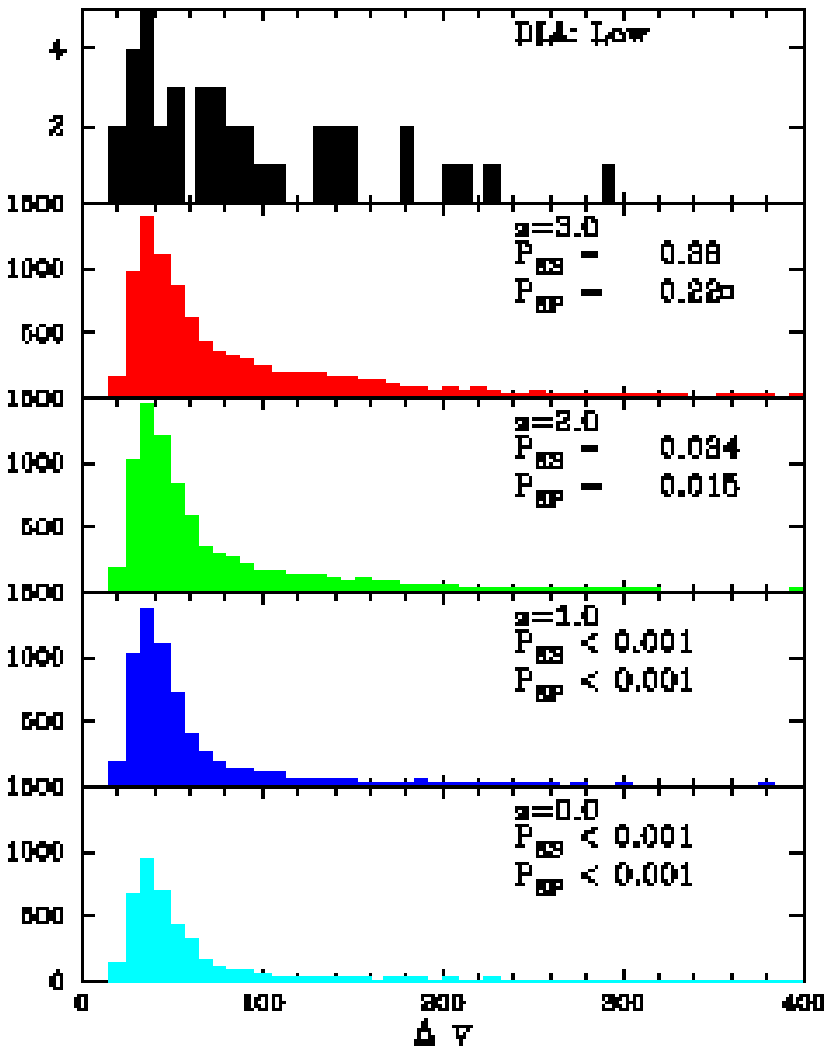}
\caption{The left panel shows the distribution of the number of galaxies (disks)
intersected by a line of sight producing a damped Lyman-$\alpha$
system, in the ``Merging Disk'' model described in the text. Note the
transition from multiple disks at high redshift, to single disks at
lower redshift. The resulting distribution of the velocity widths of
synthesized absorption line profiles is shown in the right panel. The
top panel is the data of Prochaska \& Wolfe (1998), and the remaining
panels are the models at redshifts 3, 2, 1 and 0 from top to
bottom. Note that the fraction of large $\delv$ systems
\emph{decreases} with redshift.}
\label{fig:dlas}
\end{figure}
This would appear to constitute a serious problem for hierarchical
models. However, CDM actually predicts that halos will have
considerable substructure, and Haehnelt, Steinmetz \& Rauch (1998)
found that, in their CDM-based hydro-dynamical simulations, in some
cases DLAS were produced by lines of sight intersecting multiple
``proto-galactic clumps'' orbiting within a common dark matter halo.
In these cases, the simulations reproduced the characteristic
asymmetric, edge-leading profiles, but they arose not from the
rotation of a single disk, but from circular motion with the larger
potential well of the halo. Similarly, the measured $\delv$ did not
reflect the internal rotation velocity of a single disk, but the
relative motion of the clumps within the halo. Maller et al. (2000;
MPSP) developed a ``Multiple Disk'' model, based on the same picture,
but embedded within a semi-analytic model of galaxy formation with
detailed modelling of gas cooling, star formation, and chemical
evolution.  MPSP emphasized that if the sizes of gas disks are
modelled using the standard paradigm of disk formation described
above, then the cross-section for damped absorption is dominated by
lines of sight that pass through single disks, and the previously
discussed problems apply. However, they found that a model in which
the same amount of gas was placed in more extended Mestel disks
resulted in a significant fraction of systems produced by lines of
sight passing through multiple disks, which tend to yield large
$\delv$ values. This model yielded good agreement with the data at
$z\sim2-3$ (MPSP).

In Maller et al. (in prep, see also the contribution by Maller et
al. in this volume) a more physical ``Merging Disk'' model is
explored. Using a semi-analytic model similar to the one discussed
earlier, the radial profiles of gas in isolated (non-merging) disks
are modelled using the standard approach. Merging disks are assumed to
possess more extended gas distributions (like those assumed in MPSP),
due to the gas ejected in tidal tails. In this model, because of the
strong evolution in the merger rate discussed earlier, at high
redshift ($z=2-3$), a significant fraction of the lines of sight
producing DLAS pass through two or more merging galaxies (see
figure~\ref{fig:dlas}a), and the good agreement with the kinematic
data is retained (figure~\ref{fig:dlas}b). The model makes an
interesting prediction, which is the opposite of what one would
naively expect in a hierarchical universe,
\emph{if} the $\delv$ values traced the internal rotation velocities of 
disk galaxies: in the Merging Disk model, the tail of large $\delv$
systems seen at high redshift actually diminishes at lower redshift
(fig~\ref{fig:dlas}b), where comparable observational studies have not
yet been carried out.

\section{Are the LBGs proto-disks?}
The galaxies identified at $z\ga 2$ via the Lyman-break or drop-out
technique (e.g. Steidel et al. 1999) or photometric redshifts are also
candidates for early disk galaxies. From their near-IR spectra, we
know that they have blue optical colors and show strong emission lines
--- spectrally, they would certainly be classified as very late-type
or starburst galaxies. But this does not prove that they are disks in
the true, kinematic sense --- they could be spheriods in the process
of collapsing. Morphologically, they do not really fit into any
category that can be defined based on nearby galaxies. Many have
concentrated central regions with $r^{1/4}$ light profiles, surrounded
by more diffuse exponential envelopes. Many others have peculiar
morphologies, with multiple sub-clumps and extended wispy features
reminiscent of tidal tails (Lowenthal et al. 1997). NICMOS
observations show that the objects with disturbed morphologies tend to
look similar in the near-IR (rest visual) band and in the optical
(rest-UV; Dickinson et al. 1998).

More quantitatively, Marleau \& Simard (1998) performed bulge-disk
decompositions of the light-profiles of HDF galaxies by fitting a
two-component form consisting of an exponential and S\'{e}rsic
part. The resulting bulge-to-total light ratios for galaxies with
photometric redshifts in the range $2< z < 3.5$ are shown in
figure~16g of Somerville, Faber, \& Primack (2000; SPF). About 80
percent of the HDF galaxies ($V_{606} < 26$) have at least half of
their light in a ``disk'' component by this definition, in good
agreement with the semi-analytic models.

\begin{figure}
\plottwo{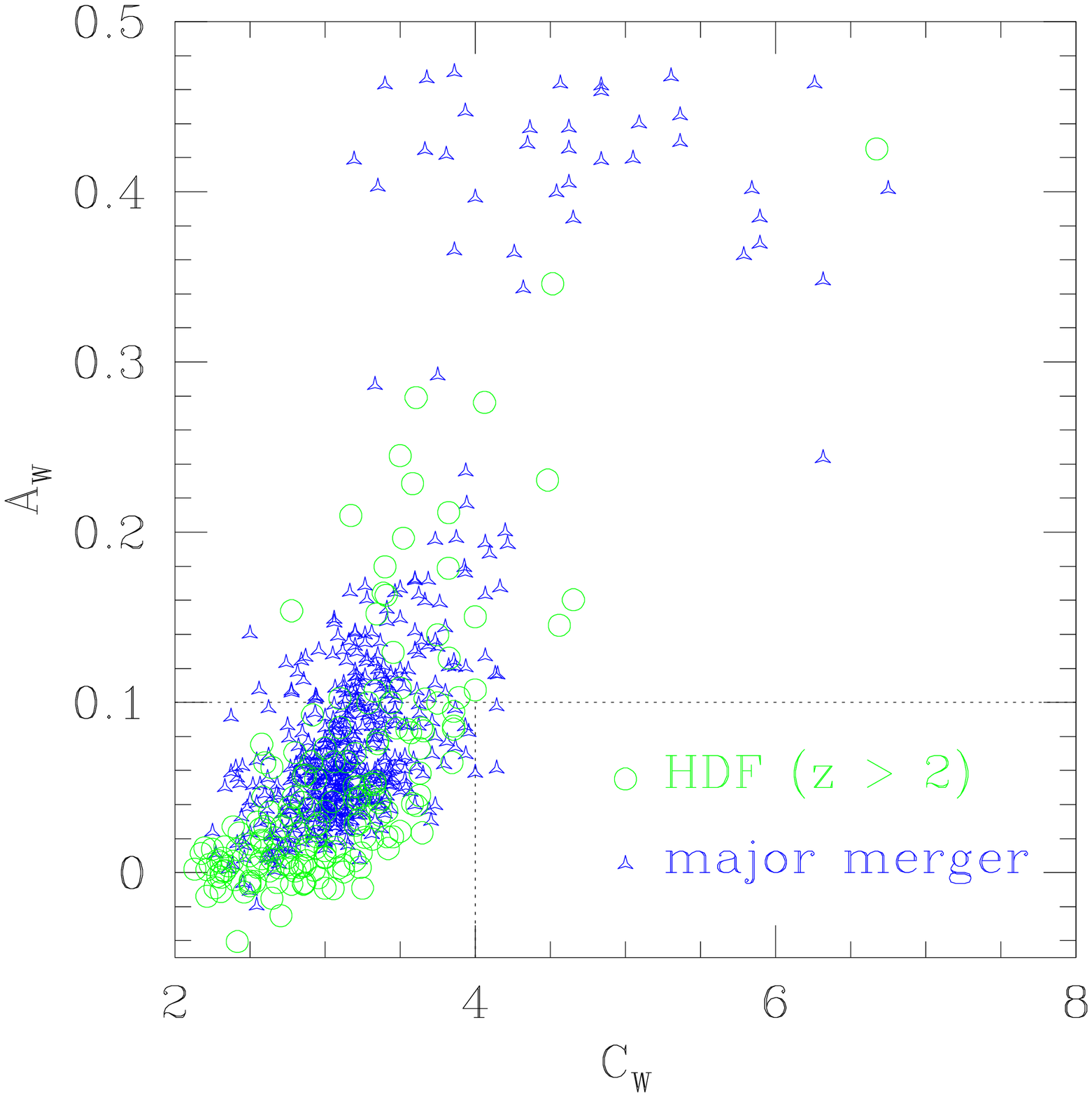}{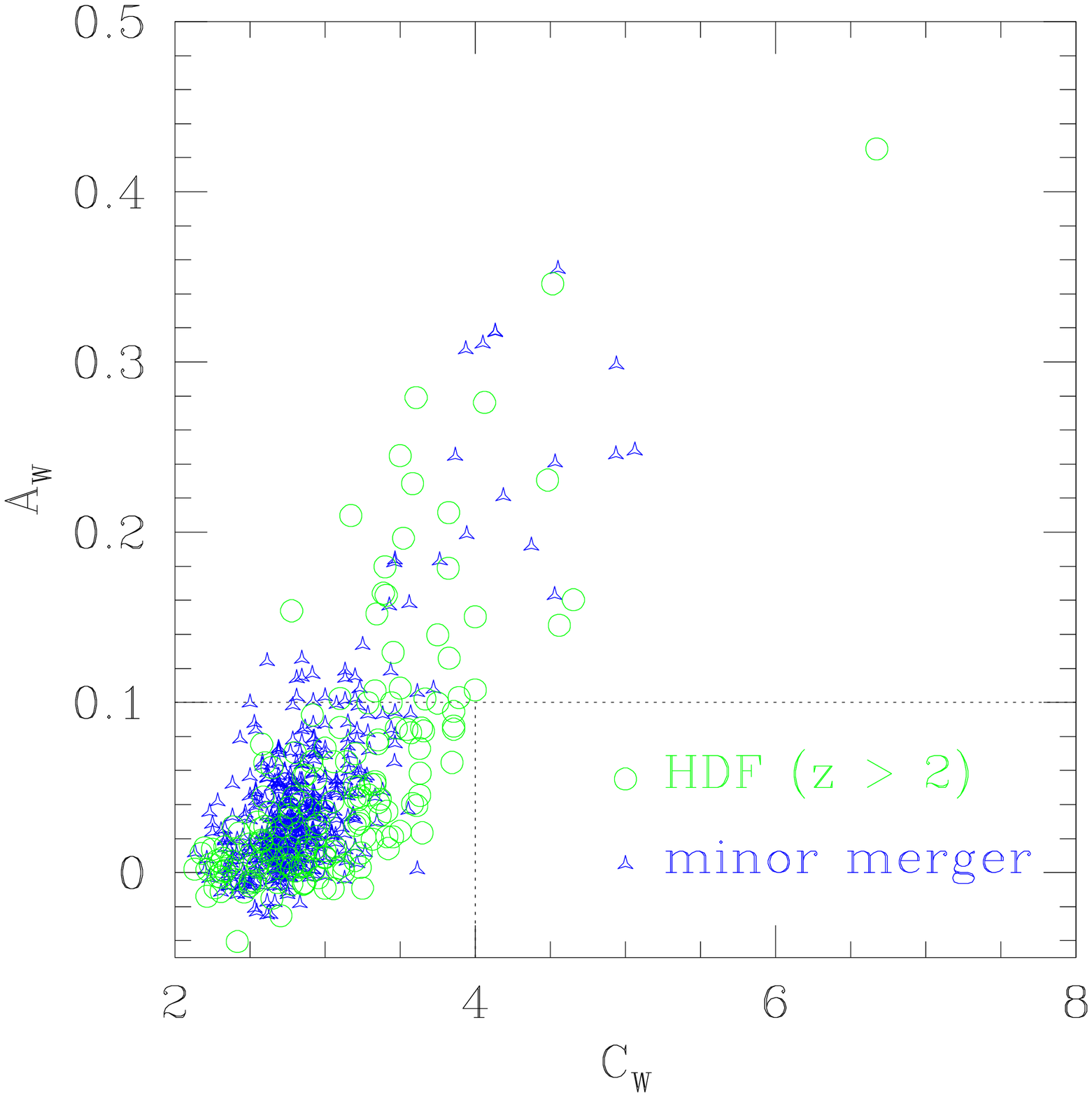}
\caption{Concentration ($C_W$) and asymmetry ($A_W$) statistics, as
defined by Wu (1999), for HDF galaxies with photometric redshifts
$z>2$ (circles) and for synthetic images from simulations of
interacting galaxies (triangles). Each simulation point represents a
snapshot in time.  Left: an equal mass merger (1:1); Right: a minor
merger (1:10). Dotted lines show the regions of the diagram inhabited
by nearby elliptical ($C_W > 4$, $A_W < 0.1$), spiral/Irr ($C_W < 4$,
$A_W < 0.1$), and interacting/peculiar galaxies ($A_W > 0.1$), from Wu
(1999). }
\label{fig:cwaw}
\end{figure}
In her thesis work, Wu (1999) performed a more detailed analysis of
the morphology of HDF galaxies. She defined concentration and
asymmetry statistics ($C_W$ and $A_W$ respectively), similar to those
defined by Abraham et al. (1996), and calibrated the new statistics on
nearby galaxies with visually typed morphologies.  In qualitative
agreement with previous studies, she found that low redshift
elliptical galaxies tend to lie in the high concentration, low
asymmetry part of the diagram ($C_W > 4$, $A_W < 0.1$), spirals in the
low concentration, low asymmetry region ($C_W < 4$, $A_W < 0.1$), and
interacting and peculiar galaxies in the high asymmetry region ($A_W >
0.1$).  When these statistics were applied to the HDF galaxies, she
found that \emph{none} of the high redshift ($z>1.2$) galaxies lie in
the part of the $C_W$-$A_W$ plane characteristic of nearby ellipticals
(see fig.~\ref{fig:cwaw}). However, many galaxies at high redshift
show high levels of asymmetry ($A_W > 0.1$), typical of local
interacting or peculiar galaxies.

We would like to use these sorts of statistics to identify galaxies
that have suffered recent mergers, and even to date the stage in the
merger and to learn something about about the type of interaction. In
order to make comparisons with theory, it is important to calibrate
the statistics against detailed simulations.  As part of an ongoing
program, we have run high resolution N-body simulations of mergers of
pairs of galaxies, including SPH treatment of gas dynamics (Hernquist
\& Katz 1989) and a
Schmidt-law recipe for star formation (Mihos \& Hernquist
1994b). Initial conditions for these interactions are set up `by
hand', choosing the properties of high-redshift disks to match the
predictions of the semi-analytic models described earlier. In work led
by P. Jonsson as part of his PhD dissertation, the star formation
history of each ``star'' particle is then convolved with stellar
population synthesis models to obtain synthetic images. We fold in the
sensitivity, resolution, and PSF of the WF/PC or NICMOS instruments
and add noise and sky background to create mock HDF images (see the
contribution of Jonsson et al., in this volume, for more details and
pretty pictures). We then run the code developed by Wu to calculate
the $C_W$ and $A_W$ statistics of our synthetic images. The results
are shown in figure~\ref{fig:cwaw} for various ``snapshots'' in time
during a major (1:1) and minor (1:10) merger, along with the results
for the HDF galaxies at $z>2$. Note that the relative weighting of the
simulated points (i.e. the number of points in a given part of the
diagram) is not meant to be comparable to the data --- we would need
to fold in the relative frequency of different types of merger and the
probability of observing a galaxy at a particular phase during the
merger. The interesting feature is the locus of $C_W$-$A_W$ space
spanned by the simulations and observations.

During much of the course of the interaction, the synthetic galaxies
inhabit the same part of the $C_W$-$A_W$ plane as nearby spirals and
irregulars (and most of the HDF galaxies). Very high levels of
asymmetry are sustained only during a relatively brief phase of the
interaction. As expected, higher levels of $A_W$ are produced by the
major merger, and are sustained for a longer fraction of the merging
time. Apparently, however, even minor mergers \emph{can} produce
significantly increased values of $A_W$ (though it is as yet unclear
how typical the case simulated, which is a prograde, coplanar merger,
will turn out to be). In fact, only two of the HDF galaxies show the
extremely high values of $A_W > 0.3$ characteristic of the most
asymmetric phase of the major merger. Most of the asymmetric HDF
galaxies ($0.1 < A_W < 0.3$) are therefore probably to be associated
with minor mergers or less asymmetric phases of major mergers. This is
perhaps to be expected, as major mergers are rare compared with minor
mergers. It is also interesting to note that several of the high
redshift HDF galaxies show the high concentrations ($C_W>4$)
characteristic of local ellipticals, but much higher asymmetries. In
the simulations, this part of the diagram corresponds to late-stage
major mergers, which will presumably become less asymmetric as they
relax, and drop down to the ``elliptical'' region of the diagram.
Though this qualitative level of agreement is encouraging, this work
should be viewed as preliminary as we have not yet considered the
effects of dust extinction, which may produce a heavily extinguished
(ULIRG-like) phase during the merger, and may significantly alter the
morphology in UV/optical bands. A broader range of parameter space
(including interactions with different orbital geometries) also needs
to be explored.

\section{Conclusions}
Observational evidence (e.g. Le F\`{e}vre et al. 2000) as well as
theoretical studies based on dissipationless N-body simulations
(Kolatt et al. 2000) support the expectation that galaxy merger rates
were higher in the past.  There is a growing realization that mergers
of gas-rich disks are instrumental in driving several important
processes: radial gas inflows, disk destruction/spheroid production
and transferring orbital angular momentum to the cold disk gas.  These
processes in turn have profound implications for several observed
populations at high redshift. Gas inflows probably fuel starbursts,
which may be responsible for activating star formation in a large
fraction of the high redshift galaxies identified in the optical and
sub-mm (SPF). At the same time, these inflows may also fuel AGN
activity (Kauffmann \& Haehnelt 2000). Producing spheroids via violent
mergers leads very naturally to the observed morphology-density
relation, as mergers take place preferentially in dense
environments. Minor mergers, which are far more common, may also
produce morphological evolution towards earlier Hubble types, by
triggering nuclear bursts of star formation. In addition, mergers may
transfer orbital angular momentum to the cold disk gas. Initially,
this gas will be launched to large distances in the form of tidal
tails, but most of it will eventually fall back to form a new
disk. This process may introduce an important new ingredient into the
`standard paradigm' of disk formation, in which all disk angular
momentum is acquired from early tidal torques.

Lines of sight passing through the extended gaseous tidal tails in gas
rich, interacting galaxies could produce the large velocity width
Damped Lyman-$\alpha$ systems observed by Prochaska \& Wolfe (1997,
1998), resolving the conflict between these observations and
hierarchical models. If correct, this picture has many interesting
implications, such as a decreasing fraction of large $\delv$ profiles
at lower redshift. If the same interaction also produces a starburst,
this also leads to specific predictions for the probability of
detecting nearby objects in emission, which we are currently
investigating (Maller et al. in prep).

The prediction of a CDM-based semi-analytic model is that many disks
exist at high redshift, but most have experienced recent interactions.
The fraction of objects showing signatures of morphological
disturbance is therefore expected to increase dramatically with
redshift. There are mounting indications that such a trend is indeed
observed. If confirmed, this constitutes direct evidence of
hierarchial structure formation in action, and may even be utilized to
obtain a measure of the merger rate as a function of redshift. One way
to quantify galaxy morphology is via statistics that simultaneously
measure the ``concentration'' and ``asymmetry'' of an image. This
appears to be a promising way to pick out interacting galaxies and
remnants of recent mergers, especially when combined with other
information such as color (Conselice et al. 2000).  In order to
interpret the observational results, however, it is important to
calibrate the statistics by calculating them in the same way on
observations and on synthetic images of simulated interacting galaxies
at various redshifts.
I have presented preliminary results from an ongoing program to carry
out such a comparison with the HDF, using the concentration and
asymmetry statistics defined by Wu (1999). An encouraging level of
agreement is obtained between the statistics calculated on the real
and synthetic images, supporting the hypothesis that the high redshift
HDF galaxies with high asymmetries are in fact ongoing interactions or
recent mergers. With further refinement of the simulations and more
extensive observations, this should be a promising way to directly
probe the merger history and assembly of galaxies in the present-day
universe.

\acknowledgements 
I thank my collaborators for allowing me to present
work in progress. I particularly thank Ariyeh Maller and Patrik
Jonsson, who actually made plots for me, and also Avishai Dekel,
Sandra Faber, Tsafrir Kolatt, Chris Mihos, Joel Primack, Jason
Prochaska, Gady Rosenfeld, Luc Simard, and Katherine Wu.

\end{document}